\documentclass{iopart}
\usepackage[utf8]{inputenc} 
\usepackage{graphicx}
\usepackage{iopams}  

\begin{document}

\title[Scalable design of an IMS cross-flow]{Scalable design of an IMS cross-flow micro-generator/ion detector}

\author{Juan J. Ortiz$^1$ , Guillermo P. Ortiz$^2$, Christian Nigri$^1$, and Carlos Lasorsa$^1$}

\address{$^1$Comisión Nacional de Energía Atómica, Av. Gral Paz 1499,
  CP 1650, San Martín, Pcia. de Buenos Aires, Argentina
  \\$^2$Dto. Física, Fac. Cs. Exactas, Nat. y A. - Inst. Mod. e
  Innov. Tec. - Universidad Nacional del Nordeste, Av. Libertad 5450
  Edificio B, W3404AAS, Campus UNNE, Corrientes, Argentina}
\ead{jjortiz@cnea.gov.ar, gortiz@exa.unne.edu.ar}
\begin{abstract}
Ion-mobility spectrometry (IMS) is an analytical technique used to
separate and identify ionized gas molecules based on their mobility in
a carrier buffer gas. Such methods come in a large variety of versions
that currently allow ion identification at and above the millimeter
scale. Here, we present a design for a cross-flow-IMS method able to
generate and detect ions at the sub-millimeter scale. We propose a
novel ion focusing strategy and tested it in a prototype device using
Nitrogen as a sample gas, and also with simulations using four
different sample gases. By introducing an original lobular ion
generation localized to a few ten of microns and substantially
simplifying the design, our device is able to keep constant laminar
flow conditions for high flow rates. In this way, it avoids the
turbulences in the gas flow, which would occur in other ion-focusing
cross-flow methods limiting their performance at the sub-millimeter
scale. Scalability of the proposed design can contribute to improve
resolving power and resolution of currently available cross-flow
methods.
\end{abstract}
\noindent{\it IMS design, gases detectors, cross-flow, micro-devices\/}
\pacs{41.75.-i, 42.82.Cr, 52.80.Hc, 85.40.-e}

\maketitle

%
%

\section{Introduction}\label{sec:intro}

Ion-mobility spectrometry (IMS) is a well established analytical
technique used to separate and identify ionized molecules in the gas
phase of a volatile compound. This technique is based on the
molecules' mobility in a carrier buffer gas. The ions acquire a drift
velocity through their mobility due to interactions with an electric
field of magnitude $E$. To the lowest order, drift velocity is
\begin{eqnarray}
v_d&=&KE  \label{eq1},
\end{eqnarray}
with $K$ the ion mobility, which allows
us to identify the compound. A dependence of $K$ with thermal fluctuation,
electrical charge, gas density and collision cross-section, can be
obtained from the balance between mobility and diffusion forces during
an elastic collision of the ionized molecule against a neutral
molecule\cite{Revercomb(1975),Instrum}

Traditional devices for time-of-flight IMS\cite{book2} come in a wide
range of sizes (often tailored for a specific application) and are
capable of operating under a broad range of conditions\cite{Trad}
above the millimeters scale\cite{trad2}. These devices use a ion
impulse field that is parallel to the flow of the carrier
gas\cite{book2}.  In contrast, aspiration condenser IMS also known as
cross-flow methods~\cite{Aspir,finger} use an impulse field that is
transverse to the flow of the carrier gas. A transverse impulse field
allows splitting of a stream of ions within the flow of the carrier
buffer gas, according to the respective mobility of
ions\cite{imsmacro}. Two main movement vectors are obtained: one in
the flow's direction and the other perpendicular to it. Devices that
apply this method are remarkably compact and relatively easy to
manufacture using micro-system technology.

There are many methods of cross flow combined with pattern recognition
and also mass-spectrometry for ion
identification\cite{Instrum,Tuovinen(2001),Utriainen(2003)}. One
problem with these method is the low resolving-power due to
overlapping of ions on detectors caused by diffusion and space charge
effects. Such effects can be reduced by increasing the flow rate as
long as laminar flow conditions remain the same. A swept-field
aspiration condenser IMS uses a variable electric drift field to move
all ion species across a single detector electrode\cite{Solis}. A
variable deflection voltage applied to a single detector electrode can
replace a detector made by an array of electrodes. An ion mobility
distribution is obtained by applying the discrete inverse Tammet
transformation to $I(V)$ data. However, reconstruction of the ion mobility
distribution is difficult when the signal to measure is comparable to
the noise and, therefore, identification possibilities of the actual
signal are reduced.

A radioactive ionization source is commonly used to produce ions.
However, in this way, ions completely cover the entrance to the
detection zone difficulting their identification. To increase
identification capabilities, some ion-focusing methods
proposed\cite{CF-teoria,CF-tor} the use of funnels to guide ions
before the splitting caused by the transverse impulse
field. Basically, ion focusing creates a concentrated starting point
for ions to start to travel forming a well-defined trajectory for each
compound and allowing to improve ion identification. Ions with
different mobility will have different
trajectories\cite{CF-teoria}. It has been utterly expensive and
complex to apply this kind of solution so far, because of the
inconveniences existing at the sub-millimeter scale. In the
ion-focusing aspiration condenser IMS, funnels or intricate channels
are implemented to narrow down the flow inside the
system\cite{CF-teoria,CF-tor}, thus generating potential turbulences
which do not allow the ions to continue their travel. This results in
a loss of identification efficiency\cite{book1} when both high flow
rates and a laminar condition are necessary.

Here, we present novel concepts for a different approach to produce
ion focusing without any physical object acting as a funnel.
\begin{figure}[t]
\begin{center}
\includegraphics[width=0.5\textwidth]{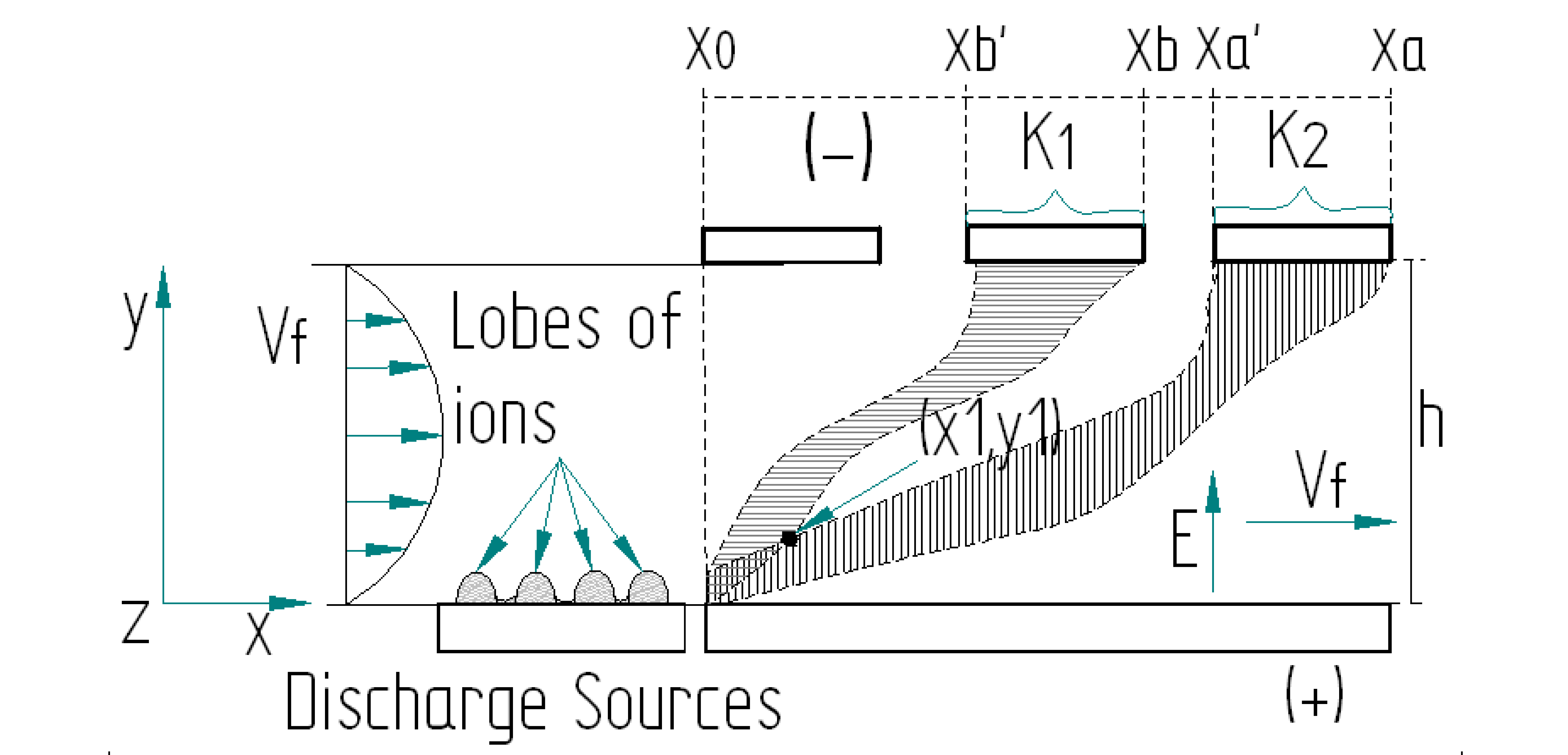}  
\end{center}
\caption{Longitudinal section of the channel with localized production
  of ions (lobes of ions) and the trajectory that follow these
  particles within the detection area. Due to ion localization,
  generated flows result in a stream of ions with small sections like
  focused ions. Under laminar flow conditions field $E$ drifts ions
  to produce the stream reaching the detectors. Ions with larger mobility
  $K_1$ reach the detectors closer to the entrance, between positions
  $(x_b,h)$ and $(x_b',h)$ whereas ions with smaller mobility $K_2$
  reaches the detectors at farther positions, between $(x_a,h)$ and
  $(x_a',h)$.}
\label{fig:separa}
\end{figure}
As illustrated in Fig.\ref{fig:separa}, ions with mobility $K_1$ and
$K_2$ , where $K_1>K_2$ starting at the same initial location near to
generated lobes of ions and defined by a stream of ions with lower
position $(x_0, 0)$ and higher position $(x_0, h_0)$ travel towards
the detection area forming two stream.  One stream reaches the
detection area that starts in $(x_b',h)$ with length $x_b-x_b'$ for
ions with $K_1$ mobility. The other stream reaches and area that
starts in $(x_a',h)$ with length $x_a-x_a'$ for ions with $K_2$
mobility. None of the ion streams will overlap on detectors, and the
required separability will be verified as long as the trajectory of
ions with larger mobility (and moving on the lowest side of the ion
stream) and the trajectory of ions with smaller mobility (moving on
the higher side of the ion stream) will meet at a point $(x_1,y_1)$
with $y_1 \leq h$, being $h$ the channel height.

The present work is organized as follows. Section \ref{sec:teoria}
presents basic equations describing the proposed model. Section
\ref{sec:dispositivo} describes the integrated prototype that we
designed containing the two main parts for ion-generation and
ion-detector. Section \ref{sec:generaion} summarizes results for
witness of corona discharge regime of the proposed
ion-generation. Section \ref{sec:crossflowfocal} describe how to deal
with the transport of micro-generated ions from their origin to
detection. Section \ref{sec:results} presents analysis of the
experimental prototype setup to generate localized ions or lobes of
ions. This section also presents a numerical simulation about the
resolving power and resolution to show the advantages of our proposed
device. Finally, Section \ref{sec:conclusion} includes our main
conclusions.

\section{Theory}\label{sec:teoria}

The hypothesis based on a localized ion generation and laminar flow
means that the access of generated ion to the detection zone is very
small, with height $y\approx0$. For a constant carrier flow in the
detection zone, two orthogonal components exist in ion movements: 
the constant flow velocity $v_f$ and the drift velocity $v_d$
Eq.(\ref{eq1}). Therefore, considering that an ion from $(x_0,0)$ uses
the same time to arrive to the detector in $(x,h)$ for each direction
of movements, we obtain:
\begin{equation}
  x-x_0=\frac{v_f\, h^2}{K\,V}\label{ec:XinvV}
\end{equation}
where the drift field can be estimated by $V/h$, assuming border
effects are negligible and electric voltage $V$, applied on the
detectors, generates a uniform field in the detection zone.  Having
defined a desired meeting point $(x_1,y_1)$ for those trajectories in
the plane $xy$ in Fig. \ref{fig:separa} and assuming uniform
velocities as in Eq.(\ref{ec:XinvV}), we can estimate
\begin{equation}
  x_1-x_0=\frac{v_f\, h}{K_1\,V}y_1=\frac{v_f\, h}{K_2\,V}(y_1-h_0).
\label{ec:foco}
 \end{equation}
Being voltage $V$ and flow velocity $v_f$ the same for both
molecules we propose that separability will be guaranteed provided
that $y_1\leq h$. Therefore, in the case $y_1=h$ we find an upper
bound for the ion-focusing size
 \begin{equation}
h_0=h (1-\frac{K_2}{K_1}), 
\label{ec:foco2}
\end{equation}
needed to separate two ionized molecules according to their mobility
and the channel height. Our main hypothesis centers on ion-focusing
being possible via the geometric design of the ion-generator.
Eq.(\ref{ec:foco2}) provides an upper bound for the scaling of this
focalization with the detection size of the device allowing
identification of the species. This requirement to scale the system
prompted us to use micro-system technology. The originality of the
present research lays in the solution reached.

\section{Description of the proposed micro-device}\label{sec:dispositivo}

\begin{figure}[h]
\begin{center}
\includegraphics[width=0.45\textwidth]{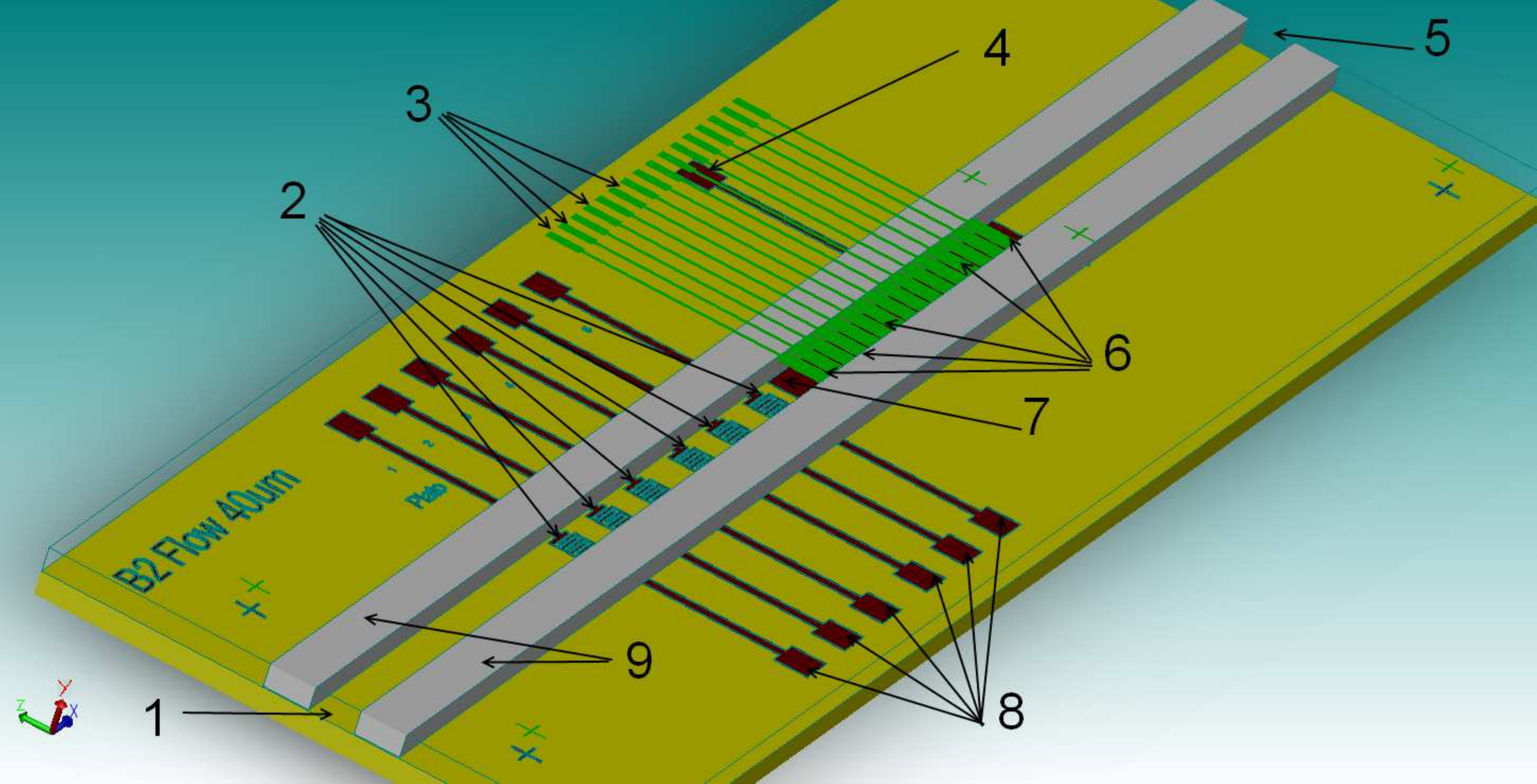}  
\end{center}
\caption{Three dimensions model of the proposed micro-device. The
  carrier gas mixes and the sample enters a guide channel determined
  by two guides (9) and two substrates (1) (only the inferior
  substrate is shown). In the area of ionization, six sources mounted
  on the surface of the inferior substrate, connect individually (2)
  through connection electrodes (8). A single connection (4) flat
  electrode (7) is mounted over inferior substrate of the detection
  zone. Sectioned flat electrodes (6) that connects individually (3)
  are mounted on the superior substrate. The carrier gas is finally
  released through an exit (5).}
\label{fig:maqueta}
\end{figure}

The device was designed considering both the position and direction of
mobility vectors of ions, from the generation and transportation of
ion up to their detection. This design was possible by arranging the
respective directions of electric discharges, fluid transportation,
and the electric fields used for detection, orthogonally to each
other. In this way, we minimized influences among the ion movements in
these directions. By employing a virtual model of the micro-device
using the CoventorWare software \cite{coventor} we constructed the
design shown in Fig.~\ref{fig:maqueta}. This software allows us to
simulate the construction of the model; from the optic layers to the
manufacturing process\cite{senturia}. With lift off, the design
proceeds; a positive photo-resin is deposited on a glass substrate and
protected with a plastic cover, containing the image of the target
structure. The resin is modified by UV light in unprotected places,
and unmodified part of the resin can be removed with the cover as
well. The remaining construct is completely covered with a layer of
copper using an anionic deposition process \cite{sputtering}. Finally,
the resin layer is removed with Acetone, leaving the geometry of the
copper layer on the substrate. This manufacturing process has been
used because of its flexibility and sturdiness. Also, one goal of this
work is obtaining a minimum size compact ion-generation/ion-detector,
with low cost and simple manufacture. 

The proposed micro-device has two main zones (see
Fig. \ref{fig:maqueta}); the ionization and the detection zone.  In
the ionization zone, lobes of ions are generated by employing an
electric discharge of corona type (see Section~\ref{sec:generaion} for
details). Charged particles of the sample are transported via a
carrier gas to the detection zone. When entering the first zone, both
the carrier gas and the sample move along the two guide through the
device towards the ionization zone. The guides determine the
separation distance between two substrates, only the inferior substrate
is shown in Fig.\ref{fig:maqueta}. The ionization zone consisting of six
sources is mounted on the surface of this inferior substrate.  Each source
has seven pairs of flat, metallic electrodes, the two electrodes;
placed in front of each other at 40$\mu$m.
The anode has six triangular tips, at a distance of 300$\mu$m from one
another, and one flat cathode. The distance between two sources is
1500$\mu$m and the last source is 540$\mu$m away from the detection
zone. These can be individually connected across the connection
electrodes, Fig. \ref{fig:maqueta} (2) and (8). 

In the detection zone, the detectors are immersed in an electrostatic
field that deflects and separates ions according to their
mobility. This zone has a flat electrode with a single connection
mounted on the inferior substrate. Individually connected and flat
sectioned electrodes are mounted on the superior substrate. Each
electrode is 900$\mu$m long and at a distance separated 100$\mu$m from
the next electrode, as shown by (6) and (3) respectively in
Fig.\ref{fig:maqueta}. The total length of the channel is
$7.5\,\,10^4\mu$m but the maximum active zone is less than half of
this length. Channel width is $2\,\,10^3\mu$m and height is
$10^3\mu$m. With these dimensions and an air stream of around
$0.33\,\,10^{14}\mu$m$^3/s$ we estimate a flow in laminar regime due
to $R_D\approx$1000.

\section{Ion micro-generation}\label{sec:generaion}

\begin{figure}[h]
\begin{center}
\includegraphics[width=0.9\textwidth]{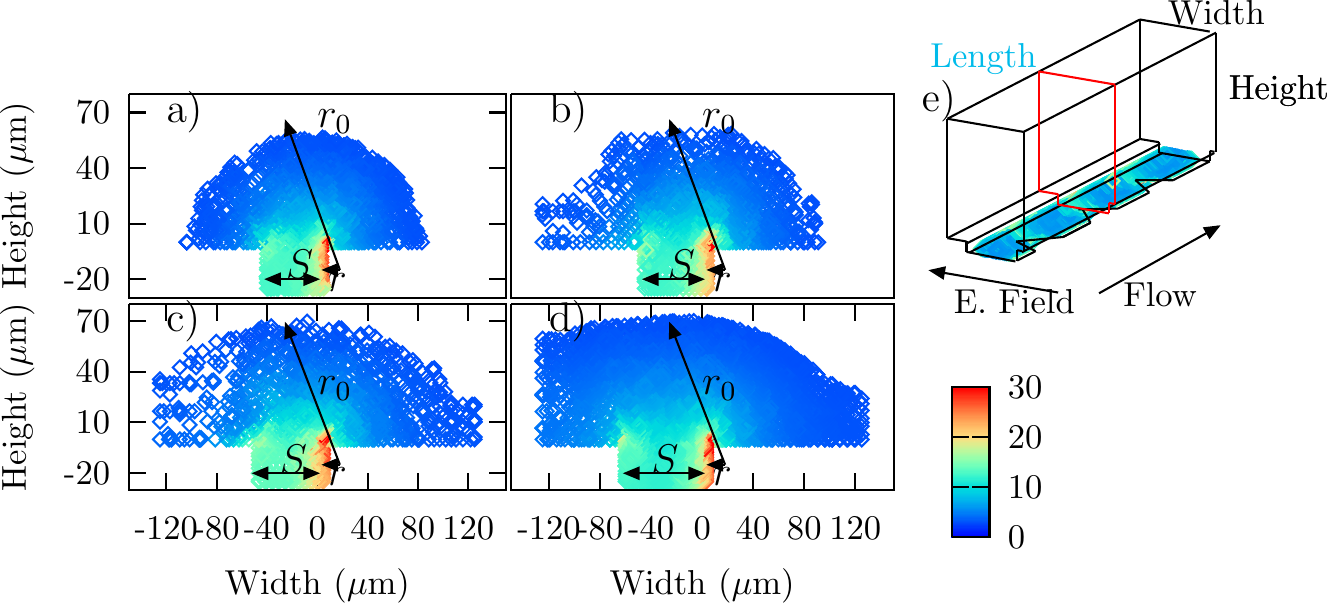}  
\end{center}
\caption{Magnitude of the electric field in V /$ \mu$m for a) $S =
  40\mu$m, $V_c = 650$ V; b) $S = 45\mu$m, $V_c = 706$ V; c)$ S =
  50\mu$m, $V_c = 807$ V; d)$ S = 60\mu$m, $V_c = 880$ V on the
  section indicated by the outline in red within the volume of the ion
  generator (e). The magnitude of the field is plotted in longitudinal
  direction on a plane 20$\mu$m above the the upper part of the anode and
  cathode. The radius of the curvature ($r = 12. 5\mu$m) of the three
  prongs forming the anode is the same for all cases in which $S$
  varies. The effective radius $r_0$ is the average distance in which
  the electric field falls to $3V/\mu$m from the anode.}
\label{fig:EsimVcExp}
\end{figure}

The ion generation problem is solved using a localized production of a
self-limited discharge. Electric discharges due to dielectric rupture
can be produced via enough energy accumulation between two electrodes
on which an electrostatic difference voltage is
applied~\cite{corona}. For our propose, the state before the beginning
of luminescence, also known as the corona discharge, is just enough to
ionize the sample. Therefore it is possible to develop the electric
field and the electric voltage before the electric discharge by
solving the Laplace equation in a volume defined by a closed
surface. In a separate work\cite{unpublished}, we study, if the
empirical Peek's law \cite{peek} that establishes the necessary
electric voltage in the anode to witness a corona discharge can be
applied at micro-scale. In summary, in these analysis, the electric
field was calculated inside the ionization volume defined in
Section~\ref{sec:dispositivo} by solving the Laplace
equations~\cite{nota} for the configuration displayed in
Fig. \ref{fig:EsimVcExp} e).
The left panel shows the magnitude of the electric field on a section
of the ion generator volume that contain the tip of the central
triangular anode. The voltage applied corresponds to critical values
experimentally determined for Nitrogen gas, $V_c=650, 706, 807$ and
$880$V and $S=40, 45, 50$ and $60\mu$m, respectively, with constant
curvature radius of the anode tip $r=12.5\mu$m. We found that it is
possible to approximately establish a same average distance $r_0$ from
the origin of the curvature radius to the surface defining a volume
where $E\geq$3 V/$\mu$m. As Peek's law is verified at the micrometer
scale\cite{unpublished}, this same average distance $r_0$ should be
approximately equal to an effective radius $r_0\approx
r+30.1\sqrt{r}$, measuring the curvature radius in micrometers. This
prediction becomes true with a 10\% sampling error in every case.

\section{Localized ions and cross-flow}\label{sec:crossflowfocal}

The proposed alternative solutions for ion focusing using a corona
discharge ion micro-generation are described in
Sections~\ref{sec:teoria}, \ref{sec:dispositivo}, and
~\ref{sec:generaion}. Considering that Eq.(\ref{eq1}) and cross-flow
design concepts can only be applied in a non-turbulent regime, we
analyzed how the fluid should travel along the channel. We propose a
channel with a rectangular section that remains constant along both
zones of the device, which means that there are no detours,
expansions, or contractions along the entire flow. Considering this,
first we verify the laminar flow condition by using the Reynold number
that we estimate for non-circular sections as
\begin{equation}
R_D=\frac{4 R_H\,v_f}{\nu}
\label{ec:Rd}
\end{equation}
in terms of the cinematic viscosity $\nu$ and the hydraulic radius
$R_H=w h/2(w+h)$ corresponding to the ratio between the fluid normal
section area and its perimeter, and $w$ as the channel width. The flow
rate $Q$ is implicit in the above equation, because it is estimated as
the average velocity of flow $v_f$ multiplied by the section of the
channel: $Q= w h v_f$.  

For the objective of our proposed design, we made a significant size
reduction for $h$ and $w$. Therefore, we need $Q$ to be low enough so
$R_D<2000$ that ensure laminar condition of the fluid regime. However,
$Q$ can not be too low, because the number of ions would be too small
to be analyzed. Also, longitudinal dimensions of the channel have to
be small enough so that the lifespan of generated ions is longer than
time they take to arrive to the detection zone. All this together
allows us to establish a higher ion velocity and a range for the
dimensions of the device. Dimensions can be optimized to obtain a
device efficiently functioning at the smallest possible scale~\cite{cf-ims}.

The flow is generated by the pressure difference between the device
entrance and exit, $P_1$ and $P_2$, respectively. Velocity is modified
by friction between the fluid and the walls due to glides of imaginary
layers of fluids. Assuming the fluid is irrotational and
incompressible and fits the device static conditions of the wall, we
analyze its dependence on vertical direction, obtaining by force
equilibrium
\begin{equation}
v_f(y)=\frac{P_1-P_2}{4 \nu L}y(h-y),
\label{ec:vf}
\end{equation}
where $L$ is the channel length and $y$ is the height variation for
the channel that goes from $0$ to $h$ as defined in
Section~\ref{sec:intro}.

\begin{figure}[t]
\begin{center}
\includegraphics[width=0.5\textwidth]{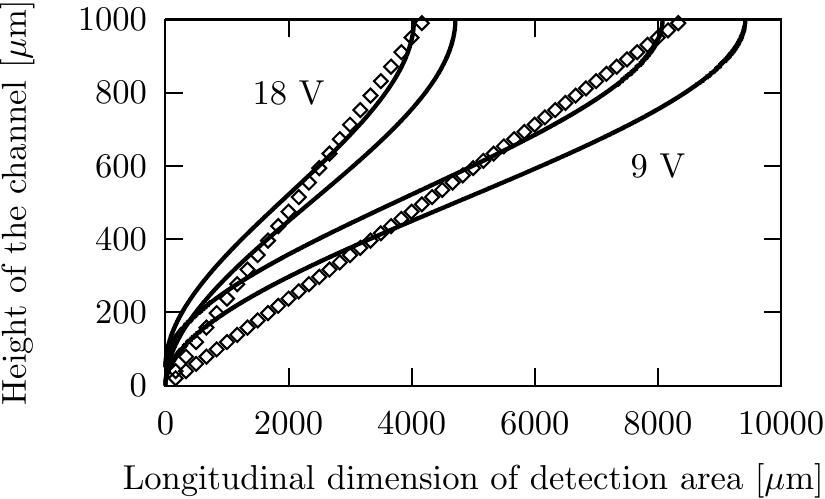}
\end{center}
\caption{Numerical simulation of trajectories for ions in the area of
  detection when applying 9 and 18$V$ voltage differences between
  cathode and anode. The continuous line for the parabolic profile
  Eqs. (\ref{ec:vf}) and (\ref{eq1}) for two initial positions (0,0)
  and (0,50$ \mu$m). Dashed dotted line illustrates a uniform
  profile with initial position (0,0).}
\label{fig:Trayec}
\end{figure}

In order to investigate the role of velocities profiles in a
sub millimeter scale, we consider numeric simulation for the
trajectory in a plane $(x,y)$ of {\em ideals} ions with the same
mobility as Nitrogen when driven in direction $x$ acquiring a
$v_f(y)$ of Eq.(\ref{ec:vf}) due to $P_1-P_2=5.35\,\,10^{-5}$ dyne /
$\mu$m$^2$, $L=7.5\,\,10^4 \mu$m and $\nu=1.822\,\, 10^{-12}$dyne
s/$\mu$m$^2$ and driven in direction $y$ acquiring $v_d$ given by
Eq.(\ref{eq1}) for $K=2.2\,\, 10^8 \mu$m$^2$/Vs due to a voltage
difference of $9$ and $18V$ between flat detector electrodes (anode
and cathode) separated $h=10^3 \mu$m. In Fig.\ref{fig:Trayec}
trajectories are shown for two initial positions (0,0) and
(0,50$\mu$m) which represent limit cases due to localizing of
ions as showed in Fig.\ref{fig:separa},
they are calculated for parabolic profile resulting a narrow laminar
stream. We also show the trajectory for uniform profile and initial
position (0,0). The trajectory with initial position (0,50$\mu$m) is
not shown, because being for the uniform profile it belongs to a
parallel line separated 50$\mu$m from the first one, defining a
straight laminar stream.

Standard cross-flow method operates at a scale of ten of millimeters
\cite{Solis}, where the change in the velocity profile due to
viscosity effects is negligible. However, at a sub-millimeter scale
viscosity effects must be taken into account. By neglecting the
velocity depends on $y$, errors in position detections as high as the
channel height $h$ could be made, as shown in
Fig.\ref{fig:Trayec}. Note that, when duplicating the applied voltage,
ions are detected at a distance that is half as long as the
corresponding one for the voltage applied originally. This is due to
inverse proportionality Eq.(\ref{ec:XinvV}) between horizontal
position of detection and voltage applied to the detector, that is
here approximately verified when considering viscosity.

\section{Results}\label{sec:results}
\begin{figure}[h]
\begin{center}
\includegraphics[width=0.55\textwidth]{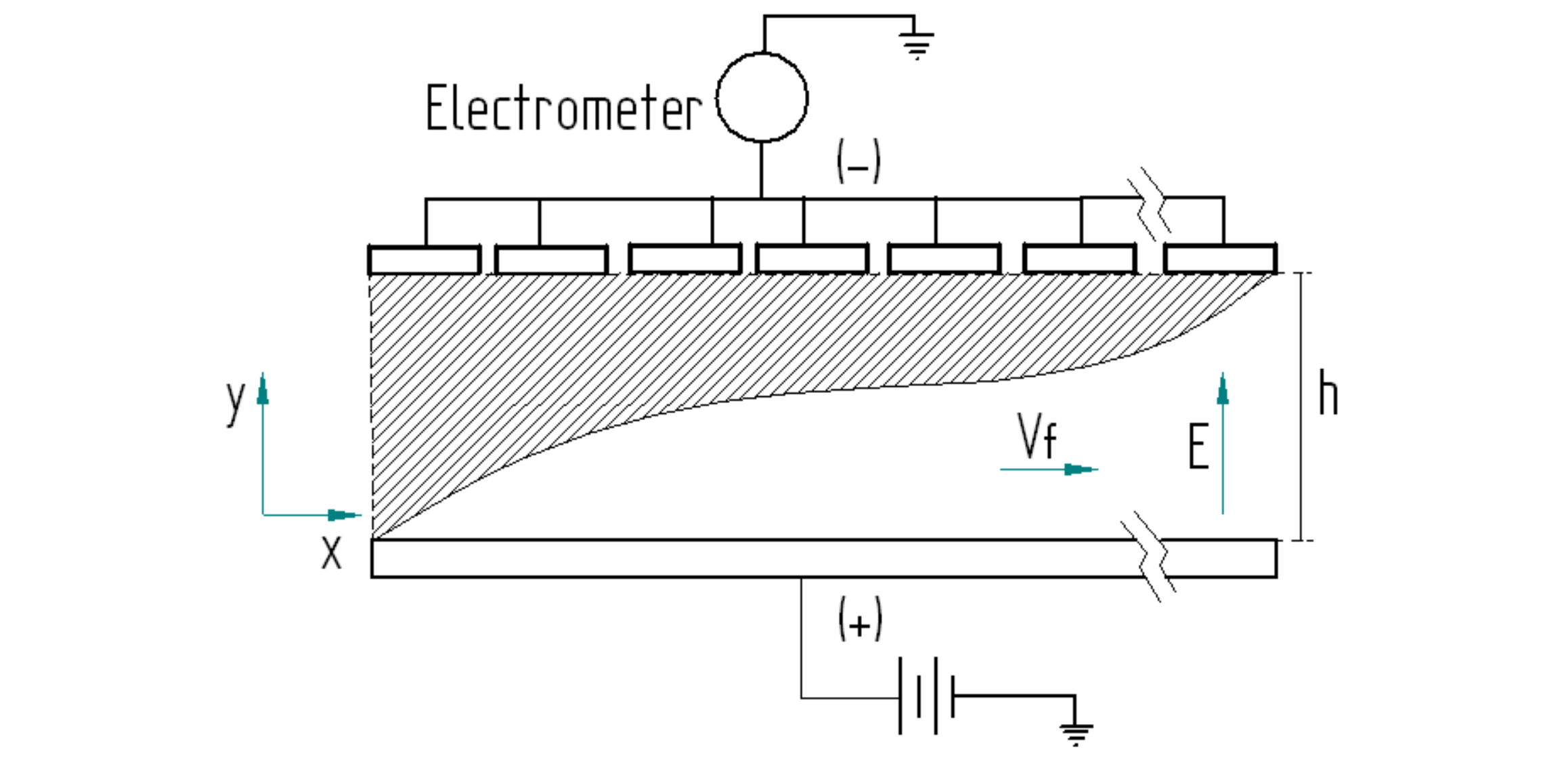}\\
\includegraphics[width=0.55\textwidth]{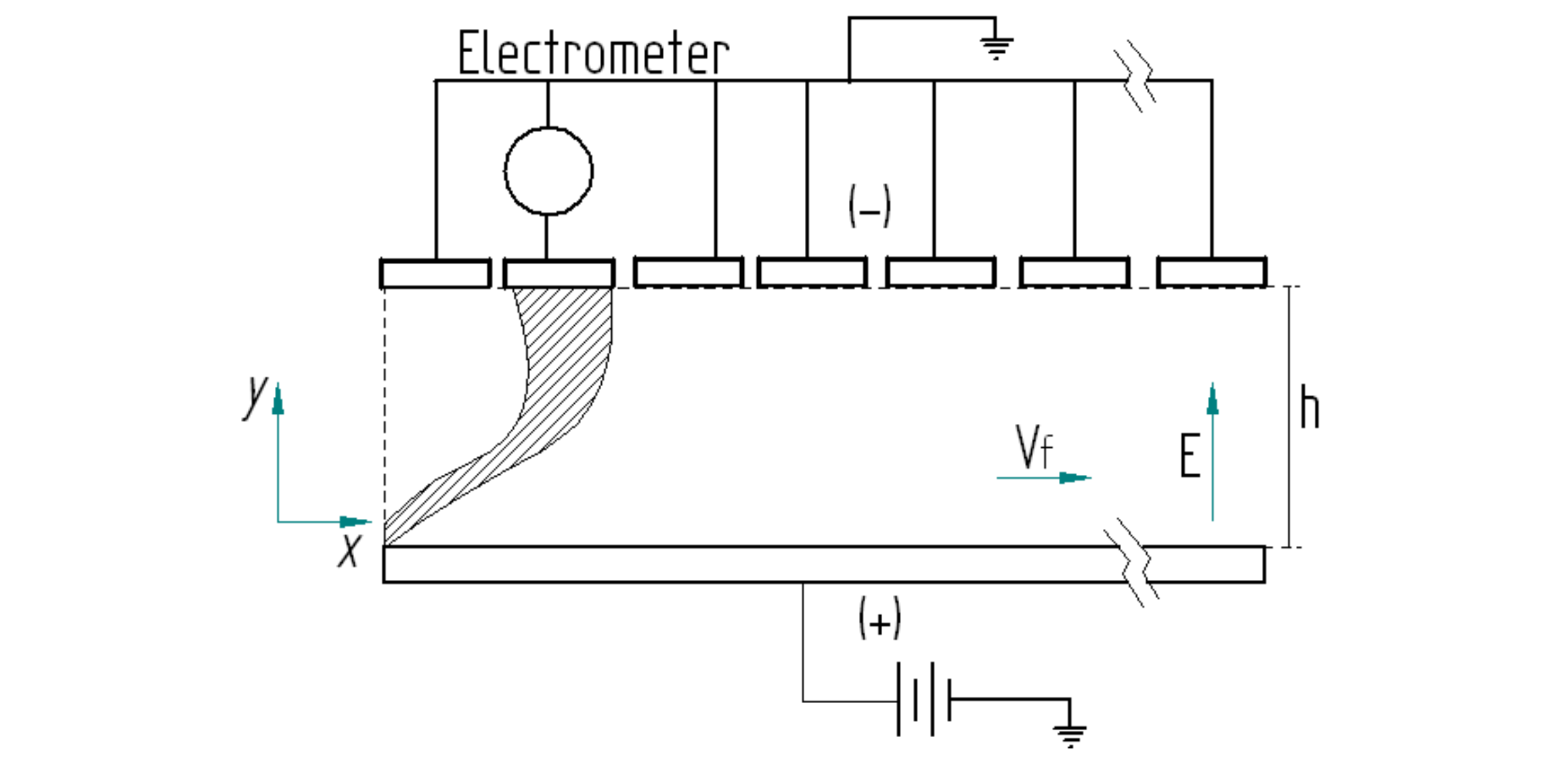}
\end{center}
\caption{\label{fig:esqMed} Charge measurements using an
  electrometer. Upper panel: measurements of the total charge as a
  function of the flow rate. Lower panel: measurements of the charge
  in each detector.}
\end{figure}

\begin{figure}
\begin{center}
\includegraphics[width=0.5\textwidth]{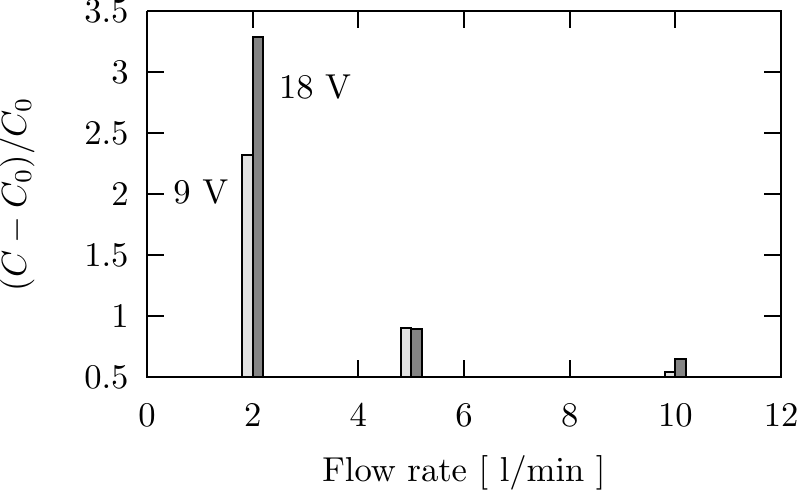}
\caption{\label{fig:vsCaudal} Charge signal $ C $ normalized to charge
  $C_0$ depending on the flow rate for $2,5,10$ $l/min$. Clear (dark)
  gray bars correspond to $9V$ ($18V$).}
\end{center}
\end{figure}

First, Nitrogen gas flows through the channel of the device,
controlled by a regulator. Then, a pulse generator is connected to a
$650V$ DC source producing every $5m$s a pulse that lasts 50$\mu$s.
We analyzed the ion charges $C$ by serially connecting a LV 8 Keithley
6514 electrometer to the arrangement of detection cathodes, as
illustrated in the upper panel of Fig.\ref{fig:esqMed}. The
accumulated charge is measured at 20s intervals.
The average is calculated over ten measures for each individual and
all electrode. Data dispersion according to the average is lower than
3~\%.  Fig.~\ref{fig:vsCaudal} shows the charge $C$ normalized with
$C_0$ depending on the flow in the device and cases in which detection
voltage is fixed at $9$ and $18V$. The reference charge $C_0$ results
from the same experiment, but using Nitrogen streaming in opposite
direction to ensures absence of ions. $C_0$ is needed because the
electrometer is sensitive enough to capture spurious signals during
the experiment, such as signals produced by generation of
ions. Parameters established in these experiments are taken from the
theoretical example analyzed in
Section~\ref{sec:crossflowfocal}. Thus, the highest charge possible is
obtained for a flow rate of 2$l/min$.  When increasing the flow rate
to 5$l/min$, the charge diminishes to more than half of the previous
charge.  Note that also the difference of $45\%$ with higher signal
for $18V$ is consistent with our theoretical model, since
concentrating detection towards half of the length of the detection
zone increases the probability to detect ions.  In the context of an
optimal design that considers the lifespan of ions, a better signal is
obtained using $18V$ rather than $9V$ because ions at $18V$ are
detected in half of the time than ions at $9V$. With flow rates of
5$l/min$ and 10$l/min$ we expect a transition towards the turbulent
regime. Because of the loss of laminar flow condition, as predicted
in Section~\ref{sec:crossflowfocal}, no significant signal differences
are expected for these flows at 9 or 18$V$.

\begin{figure}[h]
\begin{center}
\includegraphics[width=0.5\textwidth]{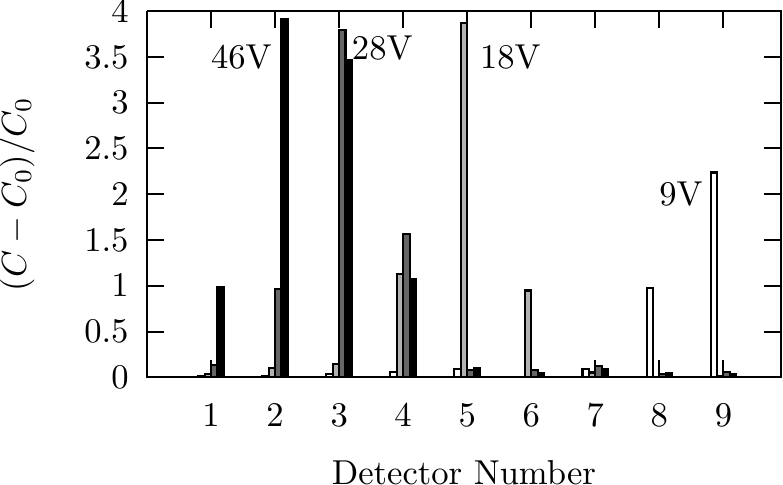} 
\caption{\label{fig:vsPotencial} Signal of charge $C$ normalized with
  $C_0$ for the flow rates 2, 5, 10 $l/min$ as function of the
  detector labels. Clear and dark gray bars correspond to 9 and 18$V$,
  respectively.}
\end{center}
\end{figure}

To verify if the generation of localized lobes of ion is enough to
separate and identify ions, the electrometer has been connected in
such a way to separately measure the signal of the charge $C$ at each
detector (see lower panel in Fig.\ref{fig:esqMed}).  We normalized the
signal with $C_0$ for a 2$l/min$ flow rate and four different
detection voltages, 9, 18, 28 and 46$V$, respectively (see
Fig.\ref{fig:vsPotencial}).
$C_0$ is obtained for each detection, by a neutral gas streaming in
the opposite direction.

When comparing results in Fig.\ref{fig:vsPotencial} with results from
the numeric simulation for detection voltages of 9 and 18$V$
(Fig.\ref{fig:Trayec}), and additionally considering the dimensions of
detectors in the micro-device (see Section~\ref{sec:dispositivo}), we
observe that the highest signal values are measured in those detectors
placed at positions and distances predicted by simulation. According
to Eq.(\ref{ec:XinvV}), there is an inverse relationship between
detection positions and applied voltages. This means that by
increasing the applied voltage two-, three-, and five-fold (from 9V to
18, $\approx$28, and $\approx$46V, respectively) the highest signal is
measured by the detector positioned at half, one third, and one fifth
of the total distance, respectively, as predicted by simulations.

This prediction is highly accurate in an ideal case of uniform
profile. In the most realistic case, a small correction is necessary,
because collection distance predictions are higher than inferred by
Eq.(\ref{ec:XinvV}) (see Section~\ref{sec:crossflowfocal}, discussion
in the last paragraph). A correct focalization by local ion
micro-generation during our experiments was crucial to experimentally
verify the model.

The proposed experimental configuration allows us to determine an
ionization volume with a diameter corresponding to as little as 10$\%$
of the channel height $h$ (see analysis in
Section~\ref{sec:generaion}).  Higher ion localizing is possible by
repeating analysis of Section~\ref{sec:generaion} and shrinking the
curvature radius of the ion generator.

With respect to separation based on ion mobility, we could state that
the performed experiments involving four voltages and one compound
could ideally be translated into an experiment involving one voltage
and four compounds. This is possible because channel height is
constant, and $E$ is approximated to $V/h$ in Eq.(\ref{eq1}), which
means that drift velocity depends only on the product of mobilities
and the applied electric voltages. Therefore, to translate results
from Fig.\ref{fig:vsPotencial} into the problem of separation by
mobility, means to consider changes in which the compounds’ mobilities
would duplicate, triplicate, and quintupling. However, expected
changes in mobility among different compounds are usually small. For
example, there is a difference of 24$\%$ between the mobility of
Toluene ($K=1.78\,\, 10^8 \mu$m$^2$/Vs) and that of Nitrogen, but for
many other compounds such differences are actually much smaller.

\begin{figure}[h]
\begin{center}
\includegraphics[width=0.65\textwidth]{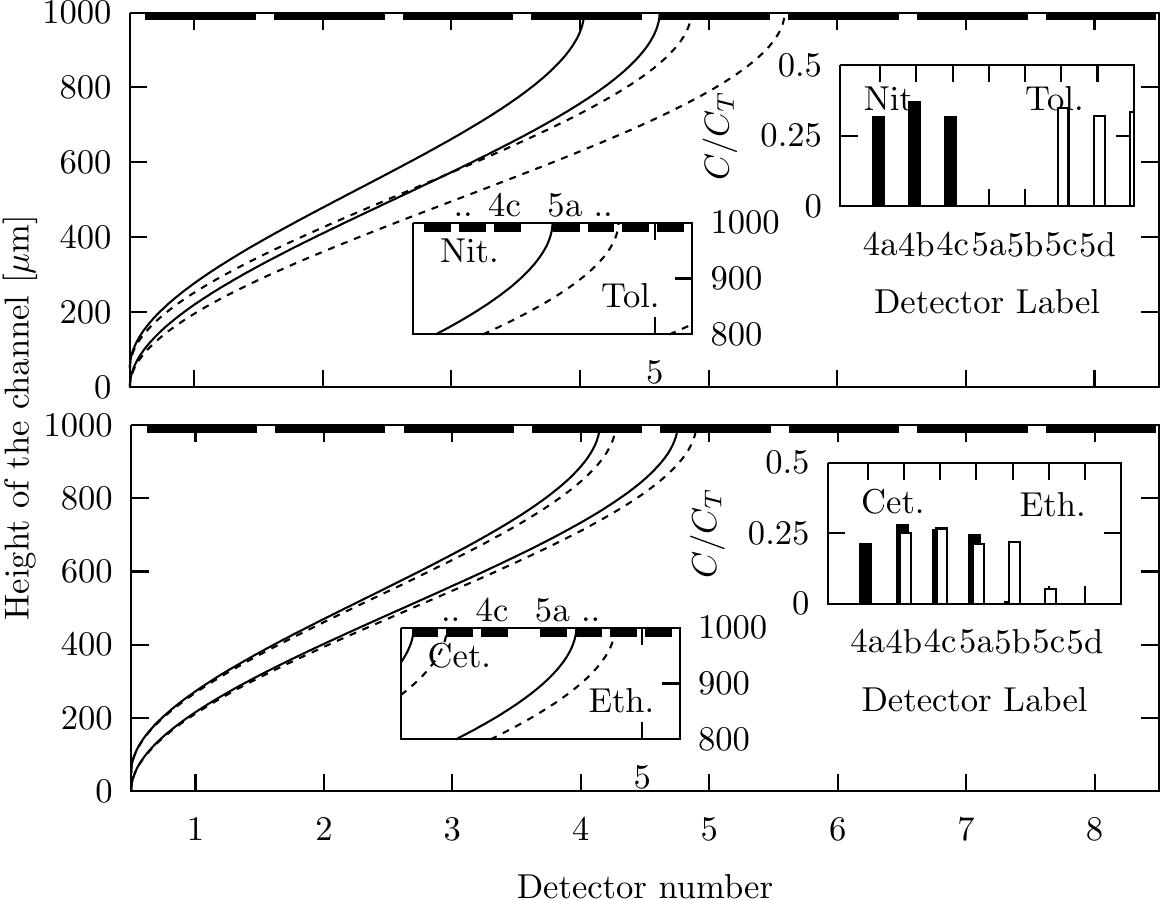} 
\caption{\label{fig:separaTNEA} Ideal trajectory of ions to the
  detectors when a voltage of 18$V$ and carrier flow rate of 2 $l/min$
  are applied to the system. The dimensions and other parameters
  correspond to real size of the device. Insets show the details where
  ions impact and normalized $C/C_T$ charge distribution on
  detectors. Segmented detector are indicated with alphanumeric label
  on each subdivision. Trajectory and IMS type spectrum for Nitrogen
  (Nit.) and Toluene (Tol.) in the upper panel, for Acetone (Cet.) and
  Ethanol (Eth.) in the lower panel.}
\end{center}
\end{figure}

Fig.\ref{fig:separaTNEA} shows numeric simulation of the ion trajectory
using a voltage of 18$V$ and a carrier flow rate of
2$l/min$.
Our design was able to separate signals corresponding to Nitrogen and
Toluene because these were not overlapping as it is shown in the
insets of upper panel Fig. \ref{fig:separaTNEA}. The insets show
details of positions where ions impact and normalized $C/C_T$ charge
respect to the total charge $C_T$ that reach detectors. Detector are
segmented and it is represented in Fig. \ref{fig:separaTNEA} with
alphanumeric labels on each subdivision. According to our analysis in
Section~\ref{sec:crossflowfocal} and employing Eq.(\ref{ec:XinvV}), we
can estimate that, for 18$V$, an increase of mobility of 24$\%$ would
be equivalent to an average displacement of an ion current from the
middle part of the fifth detector to the left by 870$\mu$m.  Because
this distance is slightly bigger than the diameter of any ion stream
reaching that position, ion streams both before and after displacement
should not overlap, thus detecting two distinct signals. This result
is consistent with our prediction about scaling between the ion
current diameter at the entrance $h_0$ of the detection zone and the
channel height $h$ (see Section~\ref{sec:teoria}). In our design,
$h_0/h\approx$ 0.1, and we verify Eq.(\ref{ec:foco2}) for Toluene and
Nitrogen. In contrast, our design was not able to separate the
overlapping signals of Acetone and Ethanol, as shown at the lower
panel of Fig.\ref{fig:separaTNEA}, because the respective mobilities
are only slightly different from one another (3.4$\%$
difference). Thus, in according to Eq.(\ref{ec:foco2}), to separate
Acetone from Ethanol, $h_0/h$ should be less than 0.034, that is not
verified with the actual size of the designed device.

The main advantage of this proposed device is its scalability that
ultimately depends on the required application. Scalability can be
tested by applying Eq.(\ref{ec:foco2}) for Ethanol and Acetone which
predicts that for channel height $h=1000\mu$m an upper bound
$h_0=34\mu$m is required for the initial diameter of the ion
stream. It is possible to separate signals of Acetone and Ethanol by
reducing the size of our proposed device by an order of magnitude,
since localization would then be smaller than the predicted upper
bound $h_0$ (see Fig.\ref{fig:separaTNEA}).
\begin{figure}[h]
\begin{center}
\includegraphics[width=0.65\textwidth]{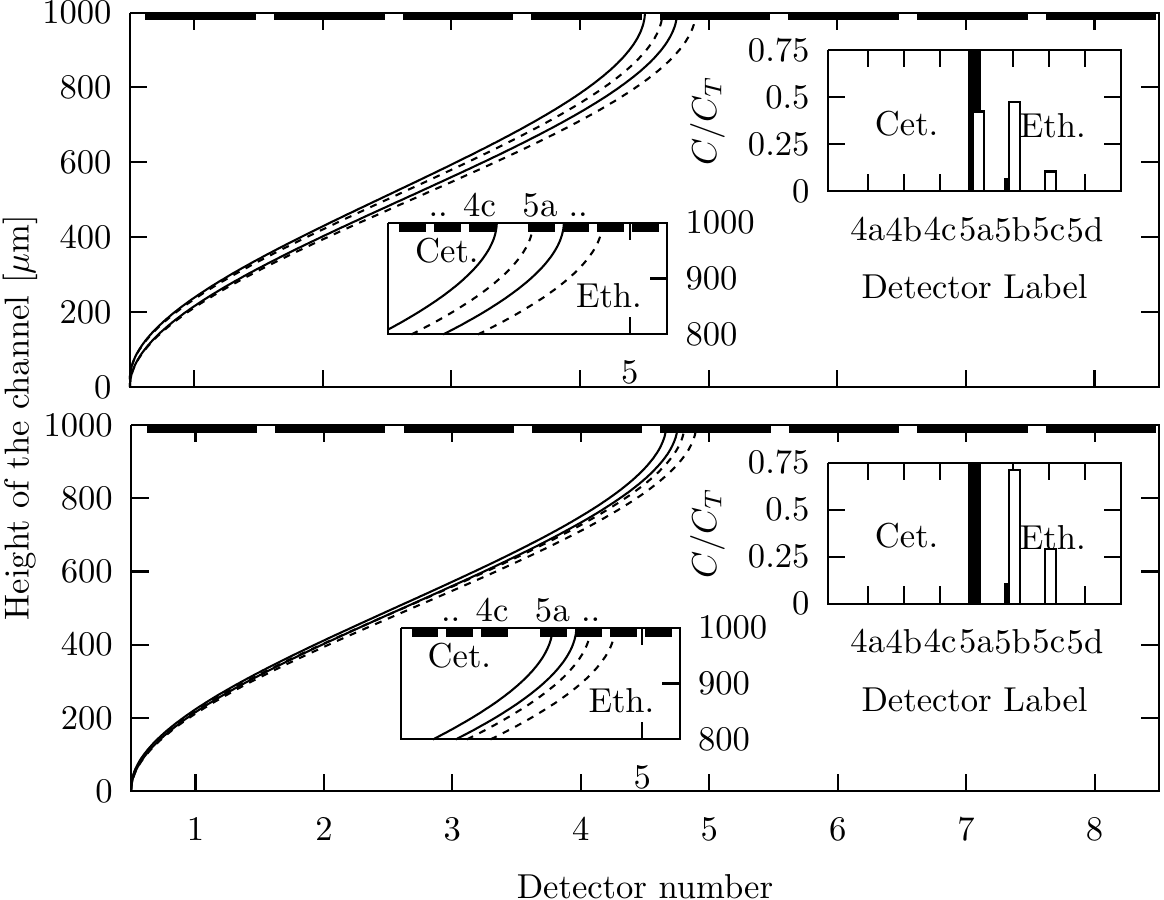}
\caption{\label{fig:separaEA} Ideal trajectory of ions of Acetone and
  Ethanol when a voltage of 180$V$ is applied. Each detector is
  100$\mu$m long. The localization of ion micro-generation is
  concentrated to $h_0=20(7)\mu$m in the upper(lower) panel. The
  carrier flow rate is 2 l/min and other parameters are as in the
  experimental device. Insets show the details where ions impact and
  the normalized $C/C_T$ charge distribution on detectors. Segmented
  detector are indicated with alphanumeric label on each subdivision.}
\end{center}
\end{figure}
The ideal trajectories of ion streams corresponding to Acetone and
Ethanol in scaled conditions with detection voltage 180 V, and
localization of ion generation smaller than 30$\mu$m, in order that
$h_0/h=0.02, 0.007$, are shown in Fig.\ref{fig:separaEA}. The other
parameters match those of the experimental device.

\begin{figure}[h]
\begin{center}
\includegraphics[width=0.5\textwidth]{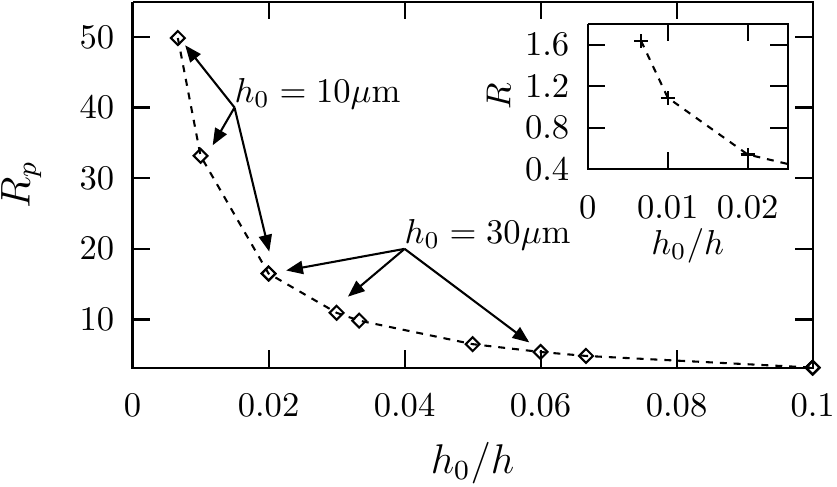}
\caption{\label{fig:respow} Resolving power $R_p$, and in the inset $R$
  to separate signals of Acetone and Ethanol as function of
  $h_0/h$. Applied voltage is adjusted according variation of $h=$500,
  1000, and 1500$\mu$m in order that $E$ is the same in all cases. The
  carrier flow rate is 2$l/min$ and other parameters are as in the
  experimental device.}
\end{center}
\end{figure}

With the proposed device the charge signal are not transformed to
average time of ion detection as it was done with other ion-focusing
method\cite{CF-tor,book1}, although the implicit ion-focusing proposed
in this work could also be adapted to be applied a Tammet
trasformation\cite{Solis}. A similar physical interpretation of
resolving power $R_p$ might be followed by defining $R_p$ as the ratio
of average trajectory position to the trajectory dispersion. Note that
with this definition does not consider the finite size of detector.
Furthermore, we are idealizing the fact that enough charge would be
detected in such cases\cite{CF-teoria}. Accordingly, the resolution
$R$ to separate two signals is defined as the ratio of the difference
between the two average trajectory position to the larger of
trajectory dispersion of both. $R_p$ and $R$ for our device to
separate signals of Acetone and Ethanol was obtained from $C/C_T$ as
it is shown in the inset of Figs. \ref{fig:separaTNEA} and
\ref{fig:separaEA} by simulations of ion trajectories using a
parabolic profile of velocity drift with random initial height of ions
starting up to $h_0$ focalization length. In Fig.\ref{fig:respow} we
show for $h=$500, 1000, 1500$\mu$m and some comparative cases of
$h_0$, the $R_p$ and $R$ as function of $h_0/h$.
We can see that $R_p$ is comparable with another ion focusing
method\cite{book1} for $h_0 = 30\mu$m and $R_p\approx10$. However, to
be able to separate Acetone and Ethanol from a mixture of the two
compounds both a higher resolving power and an increased focusing are
required. Note in the inset of Fig.\ref{fig:respow} that for
$h_0=20\mu$m and $h=1000\mu$m, $R<1$ and compounds can not be totally
separated as it is shown in upper panel of Fig.\ref{fig:separaEA},
while for $h_0=7\mu$m and $h=1000\mu$m, $R>1$ and compounds are well
separated as it is shown in the lower panel of of
Fig.\ref{fig:separaEA}. Larger $R$ imposes a challenge for flow
laminar condition. The device that we designed would be able to meet
these requirements, because it does not need any physical object that
could change fluids dynamics, as in currently available cross-flow
methods.

\section{Conclusions}\label{sec:conclusion}

In this work we address the problem to identify ion species at the
sub-millimeter scale based on their mobility and a standard cross-flow
method. To solve this problem, we propose a novel manufacturing method
using micro-system technology. We analyze the corona discharge and
verify Peek’s law in the micro-scale. With this, we dimensioned the
geometry for the proposed ion generator allowing it to localize the
ion generation and, therefore, implicitly focus ions to the entrance
of the detection zone, in a smaller size than the drift height. At the
same time, we analyzed the flow of the system to design transportation
of ions from their generation point to their detection point. By
introducing orthogonal movements we made numerical simulations to
analyze viscosity effects and state experimental configuration to
guarantee the laminar flow condition. We found a simplified design for
the construction of the micro-system, which avoids expansions or
contractions. In this way, turbulence and stopped-flow zones are
minimized in an adequate flow range, allowing us to optimize the
functioning of the device. We performed experiments on a prototype of
the proposed design to verify our hypothesis on localized ion
generation as a solution for ion species identification at
sub-millimeter scale. Variation of the average trajectory of the ion
stream with the variation of drift velocity can be analyzed with a
parabolic velocity profile. We showed that for a case in which
variations of drift velocity are 25$\%$ the charge signals displayed a
negligible overlapping of 1mm on the detection zone.  For smaller
differences in other volatile compounds we would need to adjust the
parameters used here (as these depend on the application of the
device), but not the architecture of the design. We have proposed an
upper bound for the size of the ion lobes in terms of the total drift
height and the ratio of ion mobilities to effectively separate ion
specie. Finally, we highlight how our design is elegantly simplified
compared to more complex solutions in other
methods~\cite{CF-tor,book1}. We obtained a compact micro-model that
has two advantages: it does not use radioactive sources to generate
ions, but an electrical source not exceeding 1kV, and it would reach a
higher resolving power than currently available methods.

\section*{Acknowledgments}

This study was partially supported by ANPCyT-FONCyT (grants
PICT-PRH-135-2008, PICTO-UNNE-190-2007, PAE 22594/2004 nodo NEA:23016,
and PAE-PAV 22592/2004 nodo CAC:23831). We thank Jessica Lasorsa and
Brigitte Marazzi for useful comments.


\begin{thebibliography}{20}
\bibitem{nota} The numerical calculation was done through Laplace's equation extended to the entire volume. The algorithm used is called P-FFT (precorrected fast Fourier transform) \cite {laplace1} impose conditions of symmetry on the top and sides of the volume. At the bottom edge conditions are voltage to the electrodes, positive on nails and zero for the plane, while the bottom surface charge surface boundary conditions on which find electrodes is zero \cite{coventor2}.
\end{thebibliography}
\end{document}